\documentclass[aps,12pt]{revtex4}%
\usepackage[dvips]{graphics}
\usepackage[dvips]{graphicx}%
\usepackage{amsmath}%
\usepackage{amsfonts}%
\usepackage{amssymb}
\begin{document}
\title{Very Narrow Resonances in Spherical Proton Emitting Nuclei}
\author{T.N. Leite\thanks{Present address: Colegiado de Engenharia Civil, Campus de
Ci\^encias Tecnol\'ogicas, Fund. Universidade Federal do Vale do S\~ao
Francisco, C.P. 309, 48900-000 Juazeiro, BA, Brazil. \newline Present email:
telio.leite@univasf.edu.br} and N. Teruya\thanks{nteruya@fisica.ufpb.br}}
\affiliation{Departamento de F\'{\i}sica, Universidade Federal da Para\'{\i}ba }
\affiliation{C.P. 5008, 58051-970 Jo\~{a}o Pessoa, PB, Brazil.}
\author{H. Dias}
\affiliation{Nuclear Theory and Elementary Particle Phenomenology Group, Instituto de
F\'{\i}sica, Universidade de S\~ao Paulo }
\affiliation{C.P. 66318, 05315-970 S\~ao Paulo, SP, Brazil.}

\begin{abstract}
The separation energy and half-life of some heavy proton emitting nuclei, and
the single-particle structure of the unbound $^{11}N$, have been evaluated by
implementing a careful numerical treatment to solve Schr\"{o}dinger equation
in a continuum discretization context. The basic scheme behind the method
consists in using the ground-state proton emitter in connection with an
isolated single-particle resonance. \newline{\footnotesize \textbf{Keywords:}
Decay by proton emission; lifetimes; single-particle levels.} \newline%
{\footnotesize \textbf{Journal-ref:} International Journal of Modern Physics
E, Vol. 11, No. 6 (2002) 469-473.}

\end{abstract}
\maketitle

Recently, several unstable proton emitting nuclei with medium and heavy masses
\cite{Aberg,Sellin,Davids1,Davids,Tm145,Maglione,Yu,Ta155,Davids3,Talou,Starosta}
have been discovered and are attracting much attention in both theoretical and
experimental nuclear physics. In the light mass limit unbound $^{11}N$ is
another well studied proton emitter nucleus which presents some intriguing
phenomena such as the $s_{1/2}$ intruder level and the $^{11}Be$ mirror states
discussed in the recent literature\cite{Aoyama,N11,Lep}. The parent nucleus
decays by proton emission in a quantum tunneling process. In a first
approximation we can treat this problem as an unbound $proton+core$ system in
which the ground-states instabilities are studied from the single-particle
resonance point of view. These resonances have large half-lives as long as
$\sim$ $1ns-1s$. One of the great difficulties presented in these calculations
is the determination of the separation energies and half-lives because these
states have a long lifetime in comparison with the oscillation time inside the
potential well: the energies are of the order of $\sim$ $1MeV$ and are
extremely narrow resonances $(\Gamma\lesssim10^{-12}MeV)$ corresponding to the
half-lives $t_{1/2}\gtrsim1ns$ (an enormous numerical computation time would
be necessary to sweep some $MeVs$ of energy by steps of the order of their
width size in search of the resonance position ). The spin-parity assignment
to the measured levels is another arduous task, and because the proton decay
has a strong dependence on the angular momentum, the spin-parity predictions
for some ground-state proton emitters are assigned by matching the
experimental half-life with the theoretical calculations such as the WKB
method\cite{Tm145,Ta155}. Several theoretical methods
\cite{Aberg,Maglione,Davids3,Talou} have already been presented to calculate
the half-lives and spins of these nuclei and we are presenting another method
based on solving the Schr\"{o}dinger equation in a continuum discretization
context\cite{NT,NT2}. The purpose of this work is to calculate the decay
properties, such as energies and half-lives, of some proton emitters
implementing a careful numerical treatment to this continuum projection method.

The ground-state proton emitters are treated in connection with an isolated
long lifetime single-particle resonance. Using orthogonal and complementary
projectors the scattering solution $|\psi^{+}\rangle$ of the single-particle
Hamiltonian $H$ at energy $E$ can be written as\cite{NT}%

\begin{equation}
|\psi^{+}\rangle=|u\rangle\langle u|\psi^{+}\rangle+P|\psi^{+}\rangle
,\label{Eq1}%
\end{equation}
where $|u\rangle$ is a normalized single-particle state and $P=1-|u\rangle
\langle u|$ projects onto the complementary sub-space of the single-particle
Hilbert space. Projection and formal manipulation of the Schr\"{o}dinger
equation $(E-H)|\psi^{+}\rangle=0$ then gives%

\begin{equation}
P|\psi^{+}\rangle=|\varphi^{+}\rangle+G_{PP}^{+}(E)PH|u\rangle\langle
u|\psi^{+}\rangle,\label{Eq2}%
\end{equation}

\begin{equation}
\langle u|\psi^{+}\rangle=\frac{\langle u|H|\varphi^{+}\rangle}{E-\langle
u|H|u\rangle-\langle u|HG_{PP}^{+}(E)H|u\rangle},\label{Eq3}%
\end{equation}
where $G_{pp}^{+}(E)=(E+i\eta-H_{pp})^{-1}$ is the Green's function and
$|\varphi^{+}\rangle$ is the scattering solution of the projected Hamiltonian
$H_{pp}$ at energy $E$. The $|\varphi^{+}\rangle$ solution is obtained by
solving the inhomogeneous equation%

\begin{equation}
(E-H)|\varphi^{+}\rangle=\alpha|u\rangle,\label{Eq4}%
\end{equation}
with $\alpha=-\langle u|H|\varphi^{+}\rangle$ adjusted to satisfy the
orthogonal relation $\langle u|\varphi^{+}\rangle=0$. The resonant energy,
$E_{0}$, is defined as the energy for which the integrated internal
probability density is largest and $|u\rangle$ is chosen as being proportional
to the internal part of the resonant wave function $|\psi^{+}\rangle_{E=E_{0}%
}$, truncated at a radius somewhat larger than the potential radius,
containing the principal behaviour of the resonance inside the well:%

\begin{equation}
\langle r|u\rangle=N(1+e^{(r-R_{u})/a_{u}})^{-1}\langle r|\psi^{+}%
\rangle_{E=E_{0}},\label{Eq5}%
\end{equation}
where $N$ is a normalization factor, $R_{u}$ and $a_{u}$ are the truncation
parameters for $\langle r|u\rangle$. Thus the complex single-particle
resonance energy $(\varepsilon_{u}-\frac{i\Gamma_{u}}{2}=\langle
u|H|u\rangle+\langle u|HG_{PP}^{+}(E)H|u\rangle)$ is calculated in the energy
$E_{0}$. The imaginary part $(\Gamma_{u}=2\pi|\alpha|^{2})$ corresponds to the
single-particle width and the associated half-life is given as
$t_{1/2}=\frac{\hbar}{\Gamma_{u}}\ln2$.
The potential utilized in the single-particle Hamiltonian
contains the central potential in a Woods-Saxon form and the spin-orbit term
with a Woods-Saxon derivative form, also including the centrifugal and Coulomb
(as a uniformly charged sphere) terms.

\begin{figure}[ptb]
\includegraphics[width= \textwidth]{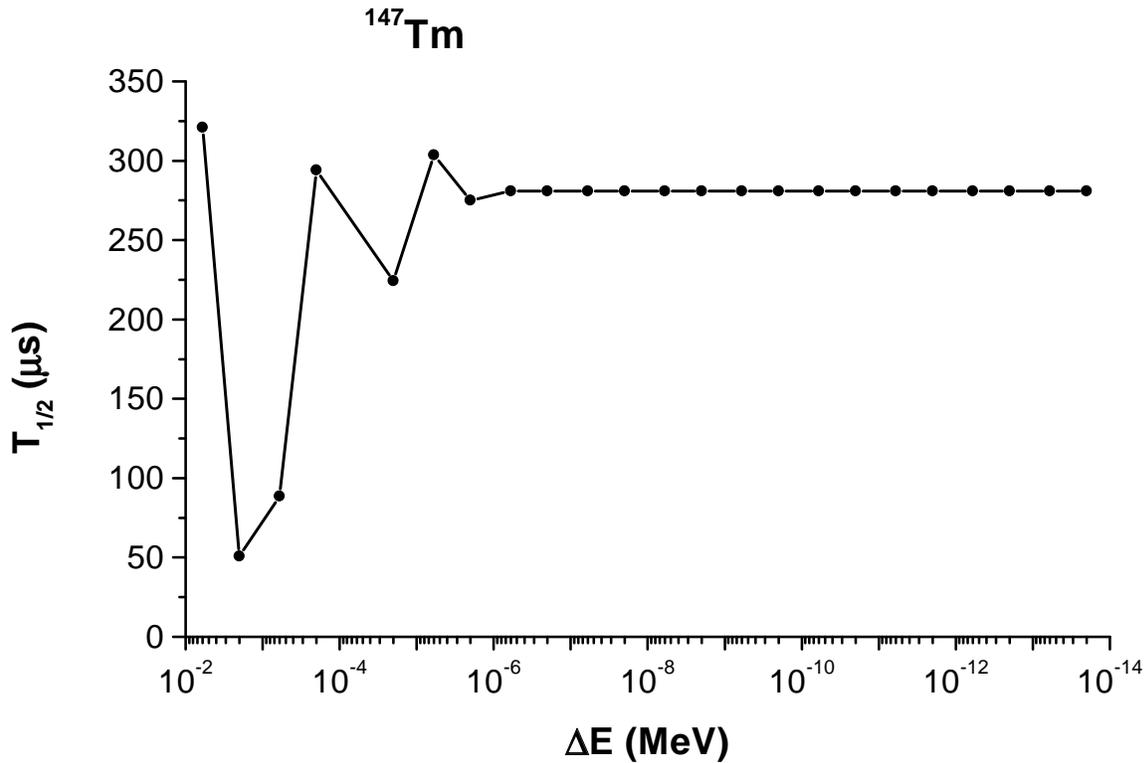}\caption{The half-life for
$^{147}Tm$ is calculated by improving the integration energy step, $\Delta E$,
that means a refinement in the energy range where the resonance is evaluated.
The results became stable when the energy precision is improved.}%
\label{fig1}%
\end{figure}

To determine the resonance parameters for the ground-state proton emitters,
special attention must be taken in the numerical implementation of this method
because these very narrow resonances are difficult to detect and require a
more accurate search. The single-particle resonance energy $\varepsilon_{u}$
is obtained imposing that it must reproduce the experimental $Q$ value by
adjusting the central and spin-orbit potentials. To avoid wasting much
computational time, and to achieve a better accuracy in the results, we start
with an initial energy range near the experimental $Q$ value to solve the
Schr\"{o}dinger equation. The energy range is swept with a integration energy
step\ $\Delta E$ to localize the energy $E_{0}$ where $\varepsilon_{u}$ and
$\Gamma_{u}$ are calculated. After this, a new energy range is defined near
$\varepsilon_{u}$. The range size is fixed at $\sim$ $2 \times10^{3}\Delta E$
centered in $\varepsilon_{u}$, and it is swept again with a energy improvement
in the integration step $\Delta E$, making a new $E_{0}$ determination in
which the $\varepsilon_{u}$ and $\Gamma_{u}$ are recalculated. This recurrent
relation is followed until the $\varepsilon_{u}$ and $\Gamma_{u}$ values
became stable. In this way we search these resonances, sweeping successive
energy ranges (sized about $2 \times10^{3}\Delta E$ centered in $\varepsilon
_{u}$), by improving the integration energy step. The Woods-Saxon central and
spin-orbit parameters (see TABLE \ref{tab0}) are adjusted by making the
stabilized energy $\varepsilon_{u}$ reproduce the experimental $Q$ value. In
FIG. \ref{fig1}, we present a typical example of this calculation for
$^{147}Tm$ for which the potential was adjusted to reproduce the experimental
ground-state energy. In TABLE \ref{tab1} our results are presented in
comparison with the other methods showing a good agreement between the data.
The small differences between our calculations and the experimental results
are due to the fact that we assume a pure single-particle configuration in a
spherical model while the real system may have soft deformations or mixing
configurations altering the occupation probabilities.
\begin{table}[ptb]
\caption{The Woods-Saxon central and spin-orbit parameters used in the
calculations. }%
\label{tab0}%
\centering
\begin{tabular}
[c]{llllllll}\hline\hline
\multicolumn{8}{l}{$V_{R}(r)=V_{0R}f(r)$ ;{\normalsize \ }$V_{ls}(r)=V_{0ls}
\left(  \frac{ h\hskip-.2em\llap{\protect\rule[1.1ex]{.325em}{.1ex}}\hskip.2em
}{m_{\pi}c}\right)  ^{2}\frac{1}{r}f^{\prime}(r) \vec{l} \cdot\vec{s}$}\\
\multicolumn{8}{l}{$f(r)=\frac{1}{1+e^{\left(  r-R\right)  /a}}$%
{\normalsize \ }; $R=1.17A^{1/3}${\normalsize \ }; $a=0.75fm$ ; $V_{0ls}%
=6.01MeV$}\\\hline
$\;\;$ & $\;\;$ & Nuclide & $V_{0R}(MeV)$ & $\;\;\;\;\;\;\;\;$ &
$\;\;\;\;\;\;\;\;$ & Nuclide & $V_{0R}(MeV)$\\\hline
$\;\;$ & $\;\;$ & $^{109}I$ & $61.699$ & $\;\;\;\;\;\;\;\;$ &
$\;\;\;\;\;\;\;\;$ & $^{155}Ta$ & $61.955$\\
$\;\;$ & $\;\;$ & $^{145}Tm$ & $63.079$ & $\;\;\;\;\;\;\;\;$ &
$\;\;\;\;\;\;\;\;$ & $^{167}Ir$ & $59.700$\\
$\;\;$ & $\;\;$ & $^{147}Tm$ & $60.398$ & $\;\;\;\;\;\;\;\;$ &
$\;\;\;\;\;\;\;\;$ & $^{185}Bi$ & $57.875$\\
$\;\;$ & $\;\;$ & $^{151}Lu$ & $63.000$ & $\;\;\;\;\;\;\;\;$ &
$\;\;\;\;\;\;\;\;$ & $\;\;\;\;\;\;\;\;$ & $\;\;\;\;\;\;\;\;$\\\hline\hline
\end{tabular}
\end{table}
\begin{table}[ptb]
\caption{The half-lives calculated in this work are compared with the other
proton emitters results. The Woods-Saxon and spin-orbit parameters are
adjusted to fit the experimental energies. }%
\label{tab1}%
\centering
\begin{tabular}
[c]{ccccccc}\hline\hline
&  &  & This work & Ref.\cite{Aberg} & Ref.\cite{Talou} & Exp.\\\hline
Nuclide & Orbit & $Q(MeV)$ & $T_{1/2}$ & $T_{1/2}$ & $T_{1/2}$ & $T_{1/2}%
$\\\hline
$^{109}I$ & $2d_{5/2}$ & $0.829$ & $7.5\mu s$ & $10\mu s$ & $-$ &
$(100\pm5)\mu s$\cite{Sellin}\\
$^{145}Tm$ & $1h_{11/2}$ & $1.728$ & $1.5\mu s$ & $-$ & $-$ & $(3.5\pm0.1)\mu
s$\cite{Tm145}\\
$^{147}Tm$ & $2d_{3/2}$ & $1.132$ & $280.0\mu s$ & $210\mu s$ & $206.8\mu s$ &
$(360\pm40)\mu s$\cite{Sellin}\\
$^{151}Lu$ & $1h_{11/2}$ & $1.255$ & $165.2ms$ & $60ms$ & $58.4ms$ &
$(130_{-50}^{+160})ms$\cite{Sellin}\\
$^{155}Ta$ & $1h_{11/2}$ & $1.776$ & $8.6\mu s$ & $-$ & $-$ & $(12_{-3}%
^{+4})\mu s$\cite{Ta155}\\
$^{167}Ir$ & $3s_{1/2}$ & $1.086$ & $193.3ms$ & $36ms$ & $-$ & $(110\pm15)ms$
\cite{Davids}\\
$^{185}Bi$ & $3s_{1/2}$ & $1.611$ & $3.6\mu s$ & $3.2\mu s$ & $3.17\mu s$ &
$(44\pm16)\mu s$\cite{Davids1}\\\hline\hline
\end{tabular}
\end{table}

In the light proton unbound nucleus $^{11}N$, the instability requires shorter
decay times than those in the heaviest ones. In this nucleus the ground-state
and the first excited state have spin-parity inversion as in its mirror
$^{11}Be$. Recent experimental data\cite{N11,Lep} report the energies and
widths of the ground and excited states in $^{11}N$. In Ref. \cite{N11} the
measured resonances are identified by adjusting the potential to reproduce the
best fit to the data. Using the same Woods-Saxon central and spin-orbit
parameters as those given in Ref. \cite{N11} we calculate the ground-state
intruder $\frac{1}{2}^{+}$ and the other excited states. The results are
showed in TABLE \ref{tab2} in comparison with the data of Refs. \cite{N11} and
\cite{Lep}. We can see that the resonant energies present good agreement with
the experimental data while the widths have small discrepancies. The major
variation in the results is due to the $\frac{1}{2}^{+}$ ground-state level
which is a broad low energy resonance. In this case, the localization of the
resonance shows up as the opposite problem to the one of the heaviest proton
emitters, because here, the resonance width is as broad as the value of its
position in energy, indicating that the half-life is approximately identical
to its short oscillation time inside the well, which brings some uncertainty
in the evaluation of the decay properties.
\begin{table}[ptb]
\caption{The calculated single-particle resonances for unbound $^{11}N$ are
presented in comparison with the experimental data. }%
\label{tab2}%
\centering
\begin{tabular}
[c]{lllllll}\hline\hline
$^{11}N$ & \multicolumn{2}{c}{This work} & \multicolumn{2}{c}{Ref.\cite{N11}}
& \multicolumn{2}{c}{Ref.\cite{Lep}}\\
$I^{\pi}$ & $E(MeV)$ & $\Gamma(MeV)$ & $E(MeV)$ & $\Gamma(MeV)$ & $E(MeV)$ &
$\Gamma(MeV)$\\\hline
$\frac{1}{2}^{+}$ & $1.38$ & $0.59$ & $1.27$ & $1.44$ & $1.63(5)$ & $0.4(1)$\\
$\frac{1}{2}^{-}$ & $2.18$ & $0.51$ & $2.01$ & $0.84$ & $2.16(5)$ &
$0.25(8)$\\
$\frac{5}{2}^{+}$ & $3.77$ & $0.47$ & $3.75$ & $0.60$ & $3.61(5)$ &
$0.50(8)$\\
$\frac{3}{2}^{+}$ & $4.81$ & $0.86$ & $4.50$ & $1.27$ & $-$ & $-$%
\\\hline\hline
\end{tabular}
\end{table}

In short, the half-lives of some proton emitters are evaluated by implementing
a careful numerical treatment to solve Schr\"{o}dinger equation in a continuum
discretization approach. The results are in good agreement with the other
theoretical and experimental data, showing that this method offers an
efficient alternative for predicting the ground and excited states in proton
emitting nuclei and these are necessary elements in more complex nuclear
structure calculations for evaluating structures other than the
single-particles ones.

\section*{Acknowledgments}

This work was supported in part by Conselho Nacional de Desenvolvimento
Cient\'{\i}fico e Tecnol\'{o}gico (CNPq), Brazil.

\end{document}